\documentclass[12pt]{article}
\title{ Gluelump spectrum in the QCD string model
}
 \author{Yu.A.Simonov\\
State Research Center\\Institute of Theoretical and
Experimental Physics, \\
Moscow, Russia}
 \date{} \newcommand{\be}{\begin{equation}}
\newcommand{\ee}{\end{equation}}

\def\fun#1#2{\lower3.6pt\vbox{\baselineskip0pt\lineskip.9pt
\ialign{$\mathsurround=0pt#1\hfil
##\hfil$\crcr#2\crcr\sim\crcr}}}

\newcommand{\ver}{\mbox{\boldmath${\rm r}$}}

\newcommand{\vep}{\mbox{\boldmath${\rm p}$}}

\newcommand{\veS}{\mbox{\boldmath${\rm S}$}}
\newcommand{\veL}{\mbox{\boldmath${\rm L}$}}

\newcommand{\vexi}{\mbox{\boldmath${\rm \xi}$}}
\newcommand{\veta}{\mbox{\boldmath${\rm \eta}$}}
\newcommand{\veB}{\mbox{\boldmath${\rm B}$}}
\newcommand{\veE}{\mbox{\boldmath${\rm E}$}}

\newcommand{\lan}{\langle}
\newcommand{\ran}{\rangle}
\begin{document}
\maketitle

\begin{abstract}

Spectrum of gluons in  the adjoint source field is computed
analytically using the
QCD string Hamiltonian, containing only one parameter -- string
tension, fixed by meson and glueball spectrum. Spin splitting is
 shown to be small.
 A good agreement is observed with spacially generated gluelump
 states measured on the lattice.
  Important role of
 gluelumps defining the behaviour of field-strength correlators is
 stressed and correspondence  with earlier computations of the latter
is established.

  \end{abstract}

\section{Introduction}

  Gluelumps \cite{1,2}  are not physical objects and their spectrum
cannot be measured on experiment. However, as will be discussed
 below, they play more fundamental role than any other hadrons since
 gluelump masses define field correlators in the QCD vacuum and in
 particular string tension. The purpose of the  present study is to
 calculate analytically the gluelamp spectrum in the framework of the
 so-called QCD string model (QCDSM). This model was developed in the
 earlier papers \cite{3,4} for spinless quarks, and was augmented by
 the spin-dependent terms treated as a perturbation \cite{5,6}.  The
 resulting Hamiltonian of the rotating string with quarks or gluons
 at the ends was recently systematically applied to mesons, hybrids
 and glueballs (see review in \cite{6} and recent developments in
 \cite{7,8}).  Three features are characteristic for the model.
 First, QCDSM is directly derived from QCD with few assumptions
 supported by the lattice data, i.e.  the minimal area law for the
 Wilson loop, and the dominant role of valence quarks and gluons.
 Second, it is fully relativistic and the Hamiltonian can be obtained
 in the c.m. system \cite{4}, or in the light--cone system \cite{9}
 or else on any other hypersurface.  Third, the model contains the
 minimal number of parameters:  fundamental string tension
 $\sigma_f=0.18 $ GeV$^2$ defined by the meson Regge slope, the
 strong coupling constant $\alpha_s$ (taken as constant in first
 approximation, $\alpha_s(\mu)$, where $\mu$ is an inverse size of
 the system, or with the freezing behaviour $\alpha_s(r)$ \cite{10}
 in more accurate calculations), and finally the  quark (antiquark)
 selfenergy $C_0=-0.25 $ GeV, which should be subtracted for each
 quark from the mass of a meson,  hybrid, or baryon. Note, that for
 gluons this term is absent because  of gauge invariance and
 therefore spectrum of glueballs and gluelumps is defined only by
 $\sigma_f$ and $\alpha_s$ .

  It is remarkable, that with these  fixed universal parameters one
  obtains spectrum of all hadronic systems in good agreement with
  lattice data and  experiment \cite{6}. In particular, the glueball
  spectrum depending only on $\sigma_f$ (with spin splittings
  depending also on $\alpha_s$)  comes  out \cite{11} in remarkable
  agreement with recent lattice data \cite{12}.

  At the same time the heavy-light mesons have been calculated in
  this method in \cite{13} including quark decay constants $f_M$,
  $ M=B,B_s, D, D_s$ again in good agreement with experiment and
  other approaches.

  This gives a hope that the QCDSM
  can be used also for the system containing a valence  gluon
  connected by an adjoint string  to the adjoint source ("infinitely
  heavy gluon") -- the so-called gluelump \cite{1,2}.

  Gluelump  has a resemblance both to  glueballs and to
  heavy-light mesons. In the framework of the string model, gluelump
  is a glueball with the  adjoint string between gluons when one
  gluon is made static, and one expects the same structure of the
  spectrum of the LS-coupling type, which is obtained in \cite{11}
  and similar   to the lattice spectrum
  calculated  in \cite{12}.

  On the other hand, gluelump differs (within the model) from
  heavy-light mesons only  in that the fundamental string is replaced
  by the adjoint  one (plus minor differences in the spin splitting
  terms  and the role  of color Coulomb forces).

  Therefore the calculation of the gluelump spectrum is expected to
  be as successful as in the case of glueballs and heavy-light mesons.
  From physical point of view the gluelumps are important at least in
  two respects. First of all gluelumps play a fundamental role in
  the nonperturbative structure of the QCD vacuum, since gluelump
  Green's functions are connected to the field correlators $\lan
  F_{\mu\nu}(x)F_{\lambda\sigma} (0)\ran$ or its invariant parts
  $D(x), D_1(x)$, introduced in \cite{14}, which are basic elements
   in the Method of Field Correlators (MFC).  In
  particular, the gluon correlation length $T_g$, defining the
  nonperturbative dynamics of confinement \cite{15,16} is simply the
  inverse mass of the lowest gluelump, and was computed in this way
  before in several systematic studies \cite{17}.

   Secondly, gluelump masses define the screening length for the
   static potential in given representation $D$ of $SU(3)$, namely
   the minimal length of the string, which may decay into two
   gluelumps.

   Recent precise measurements of static potentials $V_D(r)$ made in
   \cite{18} revealed  an accurate (within 1\%) Casimir scaling
   without any signs of  screening (string breaking) at all distances
   0.05 fm $ \leq r \leq $ 1.2 fm. To understand these results one
   should know screening length for all $D$.

   The  structure of the paper is as follows. In the first
    part of the paper the  gluelump spectrum is
   computed, with the corresponding Hamiltonian constructed in section
   2, spin  splittings considered in section 3, and spectrum
   calculation in  section 4. Comparison to the lattice data and the
   bag model is  given in section 5.
    Section 6
   contains conclusions and  outlook, while Appendix is devoted to
   the details of hyperspherical formalism  used for  calculation
   of two-gluon gluelumps.

   \section{The  gluelump Hamiltonian}

   The starting point is the gluelump Green's
   function
   $G^{glump}(x,y)$
    which is obtained from the initial and final gluelump
   operators $\Psi^{(in, out)}$, expressed in terms of the valence
   gluon field $a_\mu$ and background gluon field $B_\mu$  so that
   the total field is  $A_\mu=B_\mu+a_\mu$, with the gauge
   transformation properties
    \be B_\mu\to
   U^+(B_\mu+\frac{i}{g}\partial_\mu) U,~~ a_\mu\to U^+a_\mu U.
   \label{1}
   \ee
   Similarly to the glueball case one can write
   \be
   G^{glump}(x,y)=\lan tr_{adj}(\Gamma^{(out)}(x) G_{\mu\nu}  (x,y)
   \Gamma^{in}(y) \Phi(y,x))\ran_B,
   \label{2}
   \ee
   where $\Phi(x,y)$ is the parallel transporter representating the
   adjoint source Green's function:
   \be
   \Phi(x,y) =P\exp (ig \int^x_y\hat B_\mu(z) dz_\mu),~~ \hat B_\mu=
   B^a_\mu T^a
   \label{3}
   \ee and $G_{\mu\nu}(x,y) $ is the valence gluon Green's function
   \cite{3}
   $$
   G_{\mu\nu}(x,y) =(D^2\delta_{\mu\nu} +2ig
   F_{\mu\nu}(B))^{-1}_{x,y}=
   $$
   \be
   =\int^\infty_0 ds \int Dz e^{-K}P\exp [ ig \int^x_yB_\mu dz_\mu+
   2g\int^s_0F_{\mu\nu}(z(\tau)) d\tau]
   \label{4}
   \ee
   In (\ref{2}), $\Gamma^{(out, in)}(x)$ are operators defining
   quantum numbers of final and initial gluelump states,
   below in
   Table 1  listed are examples for the lowest mass states.

   The last term in the exponent in (\ref{4}) describes  the
   interaction of the gluon spin with the background $B_\mu$ (and
   with the adjoint string, which will be created out of it).  Here
   as well as in the glueball case \cite{11} we shall treat this
   interaction as perturbation to be discussed in section 3, and
   disregard it in the first approximation. Then  Eq. (\ref{2})
   reduces to the average of the adjoint Wilson loop with the contour
   $C$ consisting of the straight line of the source  and the path
   $z(\tau)$ of the valence gluon, integrated upon in (\ref{4}).

   Assuming the area law for the adjoint Wilson loop in agreement
   with lattice measurements \cite{18}, one can go over in a standard
   way \cite{3,4} from the Green's function $G^{(glump)}$
   to the corresponding Hamiltonian $H^{(glump)}$, connected by
   \be
   G^{glump}(x,y)
   =\lan out |\exp (-H^{(glump)} |x-y|)|in\ran.
   \label{5}
   \ee
   The resulting Hamiltonian can be obtained from that of the $q\bar
   q$ system \cite{4,6}
    when mass of one quark is zero and mass of another going
   to infinity, while the quark string tension $\sigma_f$ is replaced
   by $\sigma_{adj}=\frac94 \sigma_f$. One obtains
   $$
   H^{(glump)}= \frac{\mu}{2}+\frac{p^2_r}{2\mu}+
   \frac{L(L+1)/r^2}{2(\mu+\int^1_0 d\beta \beta^2 \nu (\beta))}
   $$
   \be
   +\frac{\sigma^2_{adj}r^2}{2} \int^1_0
   \frac{d\beta}{\nu(\beta)}+\int^1_0\frac{\nu(\beta)}{2} d\beta.
   \label{6}
   \ee
   Here $\mu$ and $\nu(\beta)$ are respectively the constituent gluon
   mass and energy density along the string coordinate $\beta$, to be
   found from minimization of the Hamiltonian.

   For the angular momentum $L=0$
   the stationary point in $\nu(\beta)$ yields a simple answer
   \be
   H^{(glump)}(L=0)=\frac{\mu}{2} +\frac{p^2_r}{2\mu} + \sigma_{adj}
   r.
   \label{7}
   \ee
   For $L>0$ instead of solving for the complicated operator
   (\ref{6}) we shall treat effect of $L$ perturbatively as in
   \cite{4}, yielding accuracy around  5\% for $L\leq 3$, namely
   \be
   H^{(glump)}(L>0)=\frac{\mu}{2} +\frac{\vep^2_r}{2\mu} +
   \sigma_{adj}  r+ \Delta M_L
   \label{8}
   \ee
   where we have defined as in \cite{4}
   \be
   \Delta M_L=-\frac{4L(L+1)\sigma^2_{adj}}{3M^3_0}
   \label{9}
   \ee
   and $\mu$ in (\ref{8}) is to be found from the minimum of the mass
   eigenvalues of $H^{(glump)}$, (accuracy of this procedure was
   found in \cite{6,19} to be better than 5\%).

   \section{Spin-dependent and color Coulomb interaction}

   We now take into account the term $F_{\mu\nu} $ in (\ref{4}),
   which  yields the gluon spin $\veS$ dependence, since
   \be
   -2i F_{ik}=2\veS\veB,~~ ik=1,2,3.
   \label{10}
   \ee
   The analysis of the spin-dependent terms can be done as in the
   case of glueballs \cite{11} and heavy quarkonia \cite{5} (note
   however, that nowhere we use the nonrelativistic inverse mass
   expansion, instead the lowest (Gaussian) field correlator $\lan
   FF\ran$ is retained in the average of (\ref{4}),  this procedure
   was tested in \cite{20} to yield accuracy around 1\%).

   As a result one obtains
   \be
   \Delta H_{LS}= \Delta H_{LS}( Thomas) +\Delta H_{LS}(pert)
   \label{11}
   \ee
   where
   \be
   \Delta H_{LS} (Thomas)=
   -\sigma_{adj}\frac{\veL\veS}{2\mu^2}\lan\frac{f(r)}{r}\ran
   \equiv b^{(Thomas)} \veL\veS
    \label{12}
    \ee
    and $f(r=\infty)=1$ takes into account that confining interaction
    starts at small $r<T_g$, as $r^2$, which significantly
    decreases the matrix element $\lan \frac{f(r)}{r}\ran$, see
    \cite{5} for details, and \cite{11} for numerical results for the
    glueball case.

    With $\Delta H(pert)$ the situation is more subtle. Indeed the
    one-gluon exchange does not contribute to the potential $V_1(r)$
    (we stick to the standard notations of Eichten-Feinberg-Gromes
    approach), and the latter is the only $LS$ piece surviving for
    the gluelump case. Hence one should consider the one-loop
    contribution, similar to the one considered in \cite{21} for heavy
    quarkonia, in the limit of one  mass infinite and another equal
    to $\mu$. In absence of actual calculations for the gluelump case,
    and to have an orientation, we shall translate the result of
    \cite{21} replacing $C_F\to C_A$, and $m_1\to \mu$,
    $m_2\to\infty$.  The result is
    \be
     \Delta H_{LS} (pert)=
   \frac{\veL\veS}{\mu^2r^3}\frac{\alpha^2_s C_A^2}{2\pi}
   (2-\ln \mu r-\gamma_E) \equiv b^{(pert)} \veL\veS .
   \label{13}
   \ee

   As we shall see the overall magnitude of $\Delta H_{LS}$ and of its
   parts is small, and finally can be neglected. For an estimate we
   shall take the brackets in (\ref{13}) to be equal to one, and the
   matrix element $\lan \frac{1}{r^3}\ran$ can be found from the
   equality  \cite{11}
   \be
   L(L+1) \lan \frac{1}{r^3}\ran= \mu\sigma_{adj}
   \label{14}
   \ee
   Consider now the contribution of perturbative gluon exchanges to
   the interaction between the valence gluon and static source.

   Here we adopt the same  attitude that was suggested and discussed
   in detail in \cite{11}. Namely, the perturbative ladder for the
   gluelump is equivalent to the BFKL ladder of the pomeron (more
   exactly to one-half of that ladder cut along the valence gluon
   line).
    For the BFKL ladder it was shown \cite{22} that the higher-order
    correction practically cancels the lowest order result. This fact
    was used in the glueball case in \cite{11} to neglect the Coulomb
    ladder completely, and  the  resulting glueball masses agree
    surprisingly well with the lattice data (whereas retaining color
    Coulomb interaction would reduce the lowest masses by 0.5 GeV,
    and increase spin splittings 3 times in strong  disagreement with
    lattice data).

    Accordingly for the gluelump case we  at first also disregard
    color Coulomb exchanges both in   mass eigenvalues and in wave
    functions,  so that relation (\ref{14}) holds true and the matrix
    element $\lan\frac{f(r)}{r}\ran$ is to be calculated with
    eigenfunctions of (\ref{7}) and (\ref{8}).  One finds that
    the spacial size of gluelump (in absence of color Coulomb force)
    is $\sqrt{2}$ times less than that of two-gluon glueballs, and
    is around 0.5 fm for lowest states. Hence the influence of
    (however reduced) Coulomb force in this system can be important.

    Therefore also calculations have been done for two other values
    of effective $\alpha_s$, which  when taking into account higher
    loop effects \cite{22} reduce to a smaller value, denoted by
    $\bar \alpha_s$. Two values of $\bar \alpha_s$, $\bar
    \alpha_s=0.15$ and $\bar \alpha_s=0.195$ have been used and the
    results for the spin-overaged mass are shown in Table 2, while
    the values of spin-orbit matrix elements $b^{(Thomas)}$ and
    $b^{(pert)}$ change for $\bar \alpha_s=0.15 $ at most 1.5 times
    and not listed in the Table.

    The resulting values of spin-splitted masses are given in Table 4
    for $\bar \alpha_s=0$.

    \section{The gluelump spectrum}

We are now in position to calculate the spectrum of (\ref{8}) and the
spin splittings (\ref{11}), (\ref{12}).

The total mass eigenvalue is written as
\be
M(J,L, n_r)= M_0(n, L) +\Delta M_L+\Delta M_{LS}
\label{15}
\ee
where $M_0(n,L)$ is the eigenvalue of the operator
$h(\mu)\equiv {\mu}/{2} +{\vep^2}/{2\mu}
+\sigma_{adj} r$, minimized over the values of $\mu$.
 \be h(\mu)
\psi(x) =\varepsilon (\mu) \psi (x). \label{16}
\ee
Using the
standard technic of the QCDSM
\cite{3,6}, one finds \be
\varepsilon(\mu) =\frac{\mu}{2} +\frac{\sigma_{adj}^{2/3} a(n,
L)}{(2\mu)^{1/3}} \label{17} \ee
where $a(n,L)$ are the eigenvalues of
the reduced equation  found numerically and given in Table 1 (lower
entries). Minimization of (\ref{17}) over $\mu$ yields for
$\mu=\mu_0$ the values given in Table 1 (upper entries). One has from
(\ref{17})
 \be
 \mu_0(n,L) = (\frac{a(n, L)}{3})^{3/4}
(2\sigma_{adj})^{1/2}.
 \label{18}
  \ee
   From $\varepsilon (\mu)$
(\ref{17}) one immediately finds that
 \be
  M_0(n,L)=2\mu_0(n,L)
\label{19}
\ee
The values of $M_0(n,L), \Delta M_L$ and the resulting spin-averaged
masses $\bar M(n,L) \equiv M_0(n,L) +\Delta  M_L
$ are given in Table 2.

Now we turn to the spin splittings of the obtained levels. The values
of $\lan \frac{1}{r^3}\ran$ are taken from (\ref{14}) and those of
$\lan\frac{f(r)}{r}\ran$ from the corresponding glueball matrix
elements in Table 7 of \cite{11}, with the proper rescaling from the
two-gluon Hamiltonian to $h(\mu)$ \cite{11}. The resulting
splittings $\Delta M_{LS}$ can be written as
\be
\Delta M_{LS} =\frac{J(J+1)-L(L+1)-2}{2} (b^{(Thomas)}(n,L) +
b^{(pert)} (n,L))
\label{20}
\ee
The values of  $b^{(Thomas)}$ and $b^{(pert)}$ are given in Table 3.
Finally summing up all corrections one obtains the resulting gluelump
masses listed in Table 4 for $\bar \alpha_s=0$.

Now we turn to the two-gluon gluelumps, which correspond to $C=+1$.
This is a special case of $3g$ glueballs when the mass of one gluon
tends to infinity. Therefore one can use the same Hamiltonian technic
as was exploited for $3g$ glueballs in \cite{11}, i.e. the
Hamiltonian $h(\mu)$ in (\ref{16}) is replaced by $h(\mu_1, \mu_2)$
(with $\mu_i$ to be found again from minimization of eigenvalues).
\be
h(\mu_1, \mu_2)=\frac{\mu_1}{2}+\frac{\mu_2}{2}+
\frac{\vep^2_1}{2\mu_1} +\frac{\vep^2_2}{2\mu_2}+
\sigma_f\left \{ |\ver_1|+|\ver_2|+|\ver_1-\ver_2|\right\}.
\label{21}
\ee

Note the appearence of $\sigma_f$ in (\ref{21}), since the string is
now fundamental and forms a triangle, passing through two gluons and
the adjoin source. The Hamiltonian (\ref{21}) can be treated using
hyperspherical formalism, for details see Appendix. Here we only
quote the result for spin-averaged masses (obtained with accuracy
better than 5\%)
\be
M(n,K)=4\mu_0+\omega_0(n+\frac12)+\frac{K(K+4)}{15} 4\mu_0
\label{22}
\ee
where $n=0,1,2,...; K=0,1,2,...$  and
\be
\mu_0=2\sqrt{\sigma_f}\left(\frac{2}{15}\right)^{1/4}
(\sqrt{2}+2)^{1/2},\omega_0=\frac{4}{\sqrt{5}}\mu_0.
\label{23}
\ee
For the lowest state, $K=0, n=0$ one obtains
\be
M^{(2g)}(0,0)=2.61 GeV (\sigma_f=0.18 GeV^2).
\label{24}
\ee
The hyperfine splitting of this level into $J^{PC}=0^{++}$ and
$2^{++}$ can be calculated in the same way as in \cite{11} with the
result that $0^{++}$ lies below $2^{++}$ with the interval
proportional to $\alpha_s$
(see Appendix for explicit expression). We place the spin-averaged
value (\ref{24}) in Table 4 and hyperfine-splitted level $0^{++}$ in
Table  5.  It is interesting to note that in two-gluon gluelumps both
hyperfine and
 tensor forces
  are operating in
   contrast to  one-gluon gluelumps, and therefore the mixing of
the state $\veB^{(1)}\veB^{(2)}$ (which is orbitally excited with
respect to $a^{(1)}_i a^{(2)}_i$) with the latter is significant,
which may explain the appearence of this state in lattice
calculations.

\section{Discussion of the spectrum. Comparison with lattice data}

The spectrum given in Tables 2 and 4  demonstrates the following
features. First of all one can see that the spin splittings of
levels are small and can be neglected in first approximation. The
ordering of levels has the same character as for two-gluon
glueball masses and can be roughly described by the equation
\be
\bar M(n,L)\cong (2n+L)\omega + \bar M_0
\label{25}
\ee
where $\bar M_0(\bar \alpha_s)\cong (1.5-2.5\bar \alpha_s)$ GeV, and
$\omega(\bar \alpha_s)\approx (0.35+\bar \alpha_s\cdot 1.2)$ GeV.
This behaviour is typical for other hadrons, see
\cite{11} for the discussion of glueballs and \cite{6} for a review
of hadronic spectra.

>From dimensional point of view the spectrum corresponds to the
increasing dimension of valence gluon operators $a_i, \partial_i
a_k,\partial_i\partial_k a_l$ etc. (and \underline{not} to the
dimension of operators $E_i, B_i, D_i B_k$  etc.).

Note that the lowest level, $1^{-- }$, is of the electric type and
there is a gap of $0.5\div 0.6 $ GeV between the ground and excited
level, while the distance between $L=0$ and $L=2$  levels is equal to
0.73 GeV $(\bar \alpha_s=0)$ or 1 GeV ($\bar \alpha_s=0.15).$

It is instructive to compare our spectrum to the gluelump spectrum of
the MIT bag model \cite{23,24}.

For $\Lambda^{1/4}=0.315$ GeV and $\alpha_s=0.23$ the masses obtained
in \cite{24} are
$$
M(1^{+-})=1.43 {\rm GeV}, M(2^{--})=1.97 {\rm GeV}, M(1^{--}) =1.98
{\rm GeV }
$$
  \be
M(3^{+-})=2.44 {\rm GeV}, M(2^{+-})=2.64  {\rm GeV}.
\label{26}
\ee

One can see the inverse ordering of the first levels in (\ref{26}),
namely magnetic level $1^{+-}$ in MIT bag model is below the electric
one $1^{--}$ and the gap between them is approximately the same as in
QCD string model, but the ordering is reversed.

Here the difference between two models is most pronounced (in
contrast  the glueball spectrum, where the lowest two-gluon
states have the same ordering in both  models, see \cite{11} and
\cite{25}). Hence the independent check of lattice calculations can
in principle decide which of the models is closer to  reality.

Now the recent lattice calculations \cite{26} (see also \cite{1,2}
for earlier lattice   studies) yield the mass   levels shown in
Table 5. One should note at this point, that the (divergent) mass
renormalization terms are not subtracted from the data and therefore
absolute scale is missing. We therefore have placed for the sake of
comparison in Table 5 the lowest $1^{+-}$ level at the same mass as
given by our calculations, see Table 2 and 4 for $\bar
\alpha_s=0.15$.

Another feature of data from \cite{26} is that only spacial links
have been used for gluelump operators at initial and final  states,
and hence the  states in Table 4 containing $E_j$ are not  directly
excited in measurements in \cite{26} (however these states could in
principle be excited indirectly due to mixing, i.e. with smaller
overlap). The states in Table 4 which are excited directly in
\cite{26} are marked by asterix and compared with lattice results in
Table 5.

One can see in Table 5 a good correspondence between the levels, and
an approximate mass degeneration of $2^{+-}, 3^{+-}$ levels is
present in both results.

However the $1^{--}, 2^{--}$ levels, almost degenerate in the QCDSM
have a gap of 0.22 GeV on the lattice. This might be explained by the
mixing of the $1^{--}$ level with the ground state, having the same
quantum numbers, whereas $2^{--}$ has no low-lying counterpart.

As was mentioned above, the crucial test of the QCDSM  is the
search for the ground state level $1^{--}$, which should lie $0.5\div
0.6$ GeV below the lowest $1^{+-}$ level found on the lattice. If
this level is not found, it would possibly mean that gluelumps are
more like bags than strings. However no derivation of the bag model
from QCD was ever made and theoretical grounds for this model are of
intuitive character, whereas the QCDSM follows  directly (with  few
assumptions supported by lattice data) from the QCD lagrangian.
Therefore the existence of the $1^{--}$ ground  state for the
gluelump seems very plausible. In conclusion of this section we
discuss   shortly an important connection between gluelumps and field
strength correlators $D(x), D_1(x)$ introduced in \cite{14} and
measured on the lattice in \cite{17} using cooling procedure and in
\cite{27} using RG smoothing. It is clear from Table 4 that two
lowest states, $1^{--}$ and $1^{+-}$, correspond to the electric and
magnetic correlators respectively, and therefore the masses
$M(1^{--})\cong 1.12$ GeV and
$M(1^{+-})\cong 1.7$ GeV  $(\bar \alpha_s=0.15)$ should give the
exponential slopes of those correlators (the notation
$D_{\parallel}$ and $D_{\perp}$ was also used in \cite{17} coiciding
with electric and magnetic respectively). The lowest mass of
$M\approx 1$ GeV is in agreement with measurements in \cite{17},
 however for $D_{\perp}$ the same slope was found in contrast with
 present calculations.  There are no general symmetry arguments,
 why masses (slopes) in both functions should be the same, (in
 contrast to electric and magnetic condensates at zero temperature)
  and this disagreement between gluelump masses and slopes in
   $D_{\parallel}, D_{\perp}$ calls for further investigation.

    One should note in addition, that the  independent calculation of
    the slope from the selfcoupled  equations for correlators
    made in \cite{28}, also gives the lowest (electric) mass
    around 1 GeV, in agreement with the present calculations for
 $(\bar \alpha_s=0.15)$.An analysis for the same quantity in the
framework of the QCD sum rules \cite{29} gives a larger value in the
quenched case, while for the magnetic mass the result was unstable.
Finally,in the lattice study \cite{30} of electric and magnetic field
correlators made without cooling or smoothing procedures,electric and
magnetic masses are obtained in good agreement with present calculations,
specifically magnetic mass turned out to be around 0.5 GeV heavier.

    \section{Conclusions}

    Results of the present calculations in the QCDSM framework are in
    general agreement with recent lattice data \cite{26}  , using the
restricted
    set of gluonic operators, and with previous lattice  and analytic
    calculations of the slope of the $D_{\parallel}$ function. It
    is not clear from the gluelump spectrum point of view why the
    correlator $D_{\perp}$ (magnetic field correlator) should have
    the same slope as that of $D_{\parallel}$,  as it was found on
    the lattice \cite{17}. With the lowest gluelump mass equal to
    $1.2 \div 1.5$ GeV in the present calculations, the onset of the
    adjoint string breaking appears at the string length $r_0=1.2\div
    1.5$ fm, in agreement with recent lattice data \cite{18}.

    The fundamental  role of gluelump states in the theory of the QCD
    vacuum makes it important to resolve the existing problem of the
    spectrum and of the relation with vacuum correlators $D_{\perp},
    D_{\parallel}$.

    The author is grateful to C.Michael for a useful correspondence,
    and to Yu.S.Kalashnikova and V.I.Shevchenko for numerous
    discussions. The present work was partially supported in the
framework of the RFFI
    projects 00-02-17836 and 96-15-96740.
    This work was completed while the author was visiting the Humboldt
    University Of Berlin in the framework of the joint DFG-RFFI project
    96-02-00088G.It is a pleasure for the author to thank M.Mueller-
    Preussker for a kind hospitality,collaboration and discussions
    and G.Bali for a series of discussions which stimulated the final
    version of the paper.

    \newpage

{\bf Table 1}\\

Valence gluon masses $\mu_0(n,L)$ (upper entries) and
reduced eigenvalues $a(n,l)$ (lower entries). $\sigma_f =
0.18$ GeV$^2, \sigma_{adj}=\frac94\sigma_f=0.405$ GeV
 \vspace{1cm}

\begin{center}
\begin{tabular}{|l|l|l|l|} \hline
$L\setminus n$&0&1&2  \\ \hline
0&0.746&1.135&1.422\\
&2.3381&4.0879&5.520\\ \hline
1&0.98&1.297&1.554 \\
 &3.3613&4.8845&6.216 \\  \hline
2&1.168&1.443&   \\
&4.248&5.63&   \\ \hline
  3& 1.33&&\\
  &5.053&&\\\hline
  \end{tabular}
\vspace{1.5cm}
\end{center}

{\bf Table 2}\\

Unperturbed masses $M_0(n,L)$, string correction $\Delta M_L$
and spin averaged masses $\bar M(n,L)$ for $L=0,1,2,3$ and
$n=0$ (upper entries and two last lines)  and
 $n=1$ (lower entries).
 \vspace{1cm}

\begin{center}
\begin{tabular}{|l|l|l|l|l|} \hline
$\setminus L$&0&1&2&3  \\ \hline
$M_0$&1.492&1.96&2.336&2.66\\
(GeV)&2.27&2.6&2.886&\\ \hline
$\Delta M_L$ &0&-0.0581&-0.103&-0.139 \\
 (GeV)&0&-0.025&-0.0546& \\  \hline
$\bar M$ &1.492&1.9&2.233&2.52   \\
(GeV)&2.27&2.575&2.83 &  \\ \hline
  $\bar M(\bar \alpha_s=0.15)$& 1.12&1.73&2.15&2.46\\\hline
  $\bar M(\bar \alpha_s=0.195)$& 0.982&1.67&2.10&\\\hline
  \end{tabular}
\vspace{1.5cm}

\end{center}
\newpage

{\bf Table 3}\\

Perturbative and Thomas spin-orbit martix elements defined
as in (\ref{11}-\ref{13})
  \vspace{1cm}

\begin{center}

\begin{tabular}{|l|l|l|l|} \hline
$L$&1&2&3  \\ \hline
$b^{(pert)}$&0.033&0.0096&0.00425\\\hline
$b^{(Thomas)}$&-0.0025&-0.0024&-0.0021\\\hline
  \end{tabular}
\vspace{1.5cm}
\end{center}

{\bf Table 4}\\

Gluelump masses for lowest states $J^{PC}$ are listed
together with gluonic  state operators in the background field
formalism and  in the general notations. Asterix marks the states
which are directly generated on the lattice \cite{26}
\vspace{1cm}

\begin{center}

\begin{tabular}{|l|l|l|c|c|l|} \hline
State&$L$&$n$&\multicolumn{2}{c|}{operator}&Mass  \\  \cline{4-5}
&&&backgr. &general&(GeV)\\ \hline
$1^{--}$&0&0&$a_i$&$E_i$&1.492\\
$1^{+-}$&1&0&$e_{ikl}\partial_ka_l$&$B_i$&1.87*\\
$2^{+-}$&1&0&$(\partial_ia_k)_{symm}$&$(D_iE_k)_{symm}$&1.93\\
$1^{--}$&0&1&$a_i$&$E_i$&2.27\\
$1^{--}$&2&0&
$(\partial_i\partial_k a_l)_J$
&$D_iB_ke_{ikl}$&2.21*\\
$2^{--}$&2&0&"&$(D_iB_k)_{symm}$&2.226*\\
$3^{--}$&2&0&"&$D_iD_kE_{l}$&2.24\\
 $2^{+-}$&3&0&$(\partial_i\partial_k\partial_la_m)_J$&$(D_iD_kB_l)_J$&2.51*\\
 $3^{+-}$&3&0&"&"&2.52*\\
 $4^{+-}$&3&0&"&$(D_iD_kD_lE_m)_4$&2.53\\
 $1^{+-}$&1&1&($\partial_ia_k)_1$&$B_i$&2.57*\\
 $2^{+-}$&1&1&($\partial_ia_k)_2$&$D_iE_k$&2.57\\
 $0^{++}2^{++}$&0&0
 &$(a^{(1)}_ia^{(2)}_k)_J$&
 $(\veB^{(1)}_i\veB^{(2)}_k+\veE^{(1)}_i\veE^{(2)}_k)_J
 $&2.61*\\\hline
  \end{tabular}
\vspace{1.5cm}
\end{center}

\newpage

{\bf Table 5}\\

The gluelump states generated by spacial operators, with masses
computed analytically in the present paper $(\bar \alpha_s=0.15)$ and
masses computed on the lattice \cite{26}. For better comparison
results of Table 2 corrected for larger $\sigma_f=0.22$ used on
lattice.
\vspace{1.5cm}

\begin{center}

\begin{tabular}{|l|l|l|l|l|l|l|} \hline
$J^{PC}$&
 $1^{+-}$&
 $1^{--}$&
 $2^{--}$&
 $2^{+-}$&
 $3^{+-}$&
 $0^{++}$\\ \hline
 $M$ (GeV)&1.87&2.23&2.45&2.84&2.84&2.96\\
 lattice &&&&&&\\ \hline
 $M$ (GeV)&1.87&2.34&2.36&2.70&2.71&2.78\\
 QCD string  &&&&&&\\ \hline
  \end{tabular}
  \end{center}

\newpage
\setcounter{equation}{0} \def\theequation{A.\arabic{equation}}

{\large{\bf  Appendix}}\\

{\bf Two-gluon gluelumps}\\

Consider a static source at point $\ver_3=0$ and two gluons at
 $\ver=\ver_1$ and $\ver=\ver_2$, connected by a fundamental string,
 passing through these three points. The Hamiltonian for the case,
 when both orbital momenta are zero (no string rotation correction)
 can be written in analogy to (\ref{7}) as
 \be
 h(\mu_1,\mu_2)=\frac{\mu_1+\mu_2}{2}+\frac{\vep^2_1}{2\mu_1}+
 \frac{\vep^2_2}{2\mu_2}
 +\sigma_f\{|\ver_1|+|\ver_2|+|\ver_1-\ver_2|\}
 \label{A.1}
 \ee
 This 3  body problem can be treated by the hyperspherical formalism
 \cite{31}. Introducing the hyperradius $\rho$ and denoting
 $\mu_1=\mu_2=\mu$, one has
 \be
 \rho^2=\veta^2+\vexi^2,~~\veta=\ver_{12}/\sqrt{2},~~
 \vexi=(\ver_1+\ver_2)\frac{1}{\sqrt{2}}
 \label{A.2}
 \ee
 one obtains the Hamiltonian for the given grand orbital momentum
  $K=0,1,2,...$
  \be
  h=+\mu-\frac{1}{2\mu\rho^5}\frac{\partial}{\partial \rho}
  (\rho^5\frac{\partial}{\partial\rho})+
  \frac{(K+\frac32)(K+\frac52)}{2\mu\rho^2}+U_K(\rho)
  \label{A.3}
  \ee
  where $U_K(\rho)$ for $K=0$ is the potential term in (\ref{A.1})
  averaged over all hyperangles. Using standard expressions from
  \cite{31} one obtains
  \be
  U_0(\rho)= C_0\sigma \rho, C_0=\frac{32\sqrt{2}(1+\sqrt{2})}{15\pi}
  \label{A.4}
  \ee
  Eq. (\ref{A.3}) can be solved numerically and results are
  tabulated, but it appears that the accuracy of about 1\% for the
  eigenvalue can be obtained by the stationary point method applied
  to the effective potential $W(\rho)$
  \be
  W(\rho)=
  \frac{(K+\frac32)(K+\frac52)}{2\mu\rho^2}+U_K(\rho);~~
  W'(\rho=\rho_0)=0;
  \label{A.5}
  \ee
  while the second derivative at $\rho=\rho_0$ defines
  the radial excitation energy $\omega$:
  \be
  W^{\prime\prime}(\rho_0)\frac{(\rho-\rho_0)^2}{2}\equiv
  \frac{\mu\omega^2(\rho-\rho_0)^2}{2}
  \label{A.6}
  \ee
  In this way one obtains the mass eigenvalue $M(\mu)$,
  and finally the stationary point condition for $\mu$,
  $\frac{dM}{d\mu}(\mu=\mu_0)=0$ defines gluon
  constituent mass $\mu_0$ and the resulting mass
  eigenvalue $M(\mu_0)$, given in (\ref{22})-(\ref{24}).
  One can note that radial excitations are given by
  $\omega_0=0.955$ GeV, and "grand-orbital excitation"
  from $K=0$ to $K=1$ is given by the interval $\Delta M
  = 0.712 $ GeV.

  The hyperfine splitting is given by the Hamiltonian
  (cf the discussion in \cite{11} for 3$g$ glueballs)
  \be
  H_{SS}= \veS^{(1)} \veS^{(2)}\frac{5\pi
  C_2(fund)\alpha_s}{3\mu_0^2} \lan
  \delta^{(3)}(\ver_{12})\ran _\rho
  \label{A.7}
  \ee
  The last factor in (\ref{A.7}) is easily computed
  \be
  \lan
  \delta^{(3)}(\ver_{12})\ran_\rho=\frac{\sqrt{2}}{\pi^2\rho^3_0},~~\rho_0
  =1.15/\mu_0.
  \label{A.8}
  \ee
  This yields the spin splitting of the levels
  \be
  \Delta M_{SS} = \veS^{(1)}\veS^{(2)} 0.49
  \mu_0\frac43\alpha_s.
  \label{A.9}
  \ee
  For $\alpha_s=\bar \alpha_s=0.15$ one obtains
  \be
  \Delta M_{SS} =0.0525 {\rm GeV} \left (
  \begin{array}{ll}
  -2,&J=0\\
  +1, & J=2
  \end{array}
  \right).
  \label{A.10}
  \ee
  In this way one gets the mass $M(0^{++})$  in Table 5.

  \end{document}